\documentclass[twocolumn]{aastex61} 
\usepackage{graphicx}
\usepackage[varg]{txfonts}
\usepackage{natbib}
\usepackage{color}

\received{}
\revised{}
\accepted{}

\submitjournal{ApJ}

\shorttitle{The radio polar brightening}
\shortauthors{Selhorst et al.}

\begin{document}

\title{Association of radio polar cap brightening with bright patches and coronal holes}

\author{Caius L. Selhorst}
\affiliation{NAT - N\'ucleo de Astrof\'isica Te\'orica, Universidade Cruzeiro do Sul, S\~ao Paulo, SP, Brazil}

\author{Paulo J. A. Sim\~oes}
\affiliation{SUPA, School of Physics and Astronomy, University of Glasgow, G12 8QQ, UK}

\author{Alexandre J. Oliveira e Silva}
\affiliation{IP\&D - Universidade do Vale do Paraíba - UNIVAP, S\~ao Jos\'e dos Campos, SP, Brazil}

\author{C. G. Gim\'enez de Castro}
\affiliation{CRAAM, Universidade Presbiteriana Mackenzie, S\~ao Paulo, SP 01302-907, Brazil}
\affiliation{IAFE, Universidad de Buenos Aires/CONICET, Buenos Aires, Argentina}

\author{Joaquim E. R. Costa}
\affiliation{CEA, Instituto Nacional de Pesquisas Espaciais, S\~ao Jos\'e dos Campos, SP,  Brazil}

\author{Adriana Valio}
\affiliation{CRAAM, Universidade Presbiteriana Mackenzie, S\~ao Paulo, SP 01302-907, Brazil}

\begin{abstract} 

Radio-bright regions near the solar poles are frequently observed in Nobeyama Radioheliograph (NoRH) maps at 17~GHz, and often in association with coronal holes. However, the origin of these polar brightening has not been established yet. We propose that small magnetic loops are the source of these bright patches, and present modeling results that reproduce the main observational characteristics of the polar brightening within coronal holes at 17~GHz. The simulations were carried out by calculating the radio emission of the small loops, {  with several temperature and density profiles, within a 2D coronal hole atmospheric model. If located at high latitudes, the size of the simulated bright patches are much smaller than the beam size and they present the instrument beam size when observed. The larger bright patches can be generated by a great number of small magnetic loops unresolved by the NoRH beam. Loop models that reproduce bright patches contain denser and hotter plasma near the upper chromosphere and lower corona. On the other hand, loops with increased plasma density and temperature only in the corona do not contribute to the emission at 17 GHz. This could explain the absence of a one-to-one association between the 17 GHz bright patches and those observed in extreme ultraviolet. Moreover, the emission arising from small magnetic loops located close to the limb may merge with the usual limb brightening profile, increasing its brightness temperature and width.}

\end{abstract}


\section{Introduction}
 {Bright areas in the polar regions of the Sun have been frequently reported at radio  {to infrared} frequencies, ranging from $\sim$15~GHz to 860~GHz \cite[see  ][and references therein]{Selhorst2003}.} {Most of those observations} were obtained by  {single-dish}  {telescopes} with low spatial resolution,  {posing challenges to draw firm conclusions about the physical origin of the increase in emission}.

\cite{Efanov1980} observed the presence of  {bright regions near the poles}
 {at 22 and 37~GHz} during the  {period of} minimum solar activity,  {and reported that such bright regions were not seen during the maximum of solar activity.} 

 Similar  {findings} were also reported  {through} other  {single-dish} observations \citep{Riehokainen1998,Riehokainen2001},  {also} suggesting that the polar brightening {could be associated with} regions in which the white-light polar faculae are observed and follows their cycle, i.e., anti-correlated with the solar cycle.  

 {Great advances} in the study of {the polar} brightening were obtained due to interferometric solar observations at 17~GHz by the Nobeyama Radioheliograph \citep[NoRH, ][]{Nakajima1994}, in operation since 1992. \cite{Shibasaki1998} concluded that these {polar cap bright regions} observed at 17~GHz was the sum of two components:  {a limb brightening effect superposed on bright features intrinsic to the poles,}  that can increase the brightening up to 40\% above the quiet Sun temperature. 

The polar  {cap} brightening at 17~GHz is characterised by the presence of small bright structures (bright patches) that appear in the regions close to the limb, with their sizes ranging from the NoRH beam size (about $15''$) up to $50''-55''$ \citep{Nindos1999}. Through synoptic limb charts, \cite{Oliveira2016}  showed a good association between the presence of coronal holes and the 17~GHz polar brightening in the period of 2010-2015. Moreover, the authors attributed the enhancement of radio brightness in coronal holes to the presence of bright patches closely associated with the presence of intense unipolar magnetic fields. 

\begin{figure}[h]
\centerline{ {\includegraphics[width=9cm]{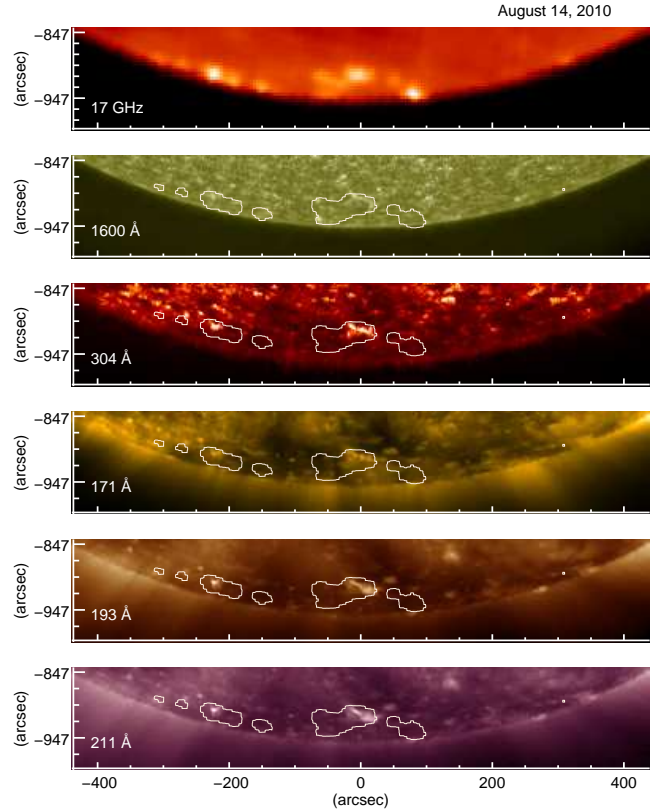}}}
\caption{Comparison between the 17~GHz polar bright patches and the EUV images obtained by SDO/AIA. The 17~GHZ contour curves correspond to 15\% above the quiet Sun temperature.}
\label{fig:obs}
\end{figure}

{  In Figure~\ref{fig:obs}, we present en example of 17~GHz bright patches (top panel), with a good correspondence with small bright structures observed in extreme ultraviolet (EUV) emission, from images of the Atmospheric Imaging Assembly (AIA) instrument \citep{Lemen2012}, on board of the Solar Dynamics Observatory (SDO). Moreover, the EUV lines formed above the transition region (171, 193 and 211~\AA) show that these bright structures are embedded in a coronal hole. Nevertheless, not all bright structures observed in EUV have 17~GHz association, as reported before \citep{Nindos1999,Riehokainen2001,Nitta2014}. }

Apart from being more frequently observed at the poles, the association between coronal holes and the presence of bright patches was also observed at lower latitudes \citep{Gopal1999,Maksimov2006}. 

 While the radio limb brightening is now relatively well understood \citep{Selhorst2005a}, the origin of the intrinsic bright patches near the solar poles has not been identified yet. In this work, we propose a model to explain the observed radio bright patches within coronal holes near the solar poles. Using small magnetic loop models to represent the source of the bright patches, we were able to reproduce the typical brightness temperature and size of the small (around 10'') polar bright patches. We suggest that larger regions ($\sim 50$'') are formed by a number of small loops, unresolved by NoRH.

\section{Modeling coronal holes and polar bright patches}

In this section, we describe our proposed atmospheric model for coronal holes and the small magnetic structures to represent the origin of the polar bright patches.

\subsection{The atmospheric model}

\cite{Selhorst2005a} proposed an atmospheric model  {(hereafter referred to as the SSC model)} with the distributions of temperature and density (electron and proton) as a function of height, from the photosphere up to $40,000$~km in the corona. {  To calculate the 17~GHz limb brightening and verify the influence of spicules, the radiative transfer was performed through a 2D space, in order to account for the curvature of the Sun, from the disc center to the limb, and the SSC solar atmosphere. In this work, we follow the same procedure, with the appropriate atmospheric model for coronal holes (Section \ref{sec:ch}) and inclusion of magnetic loops to represent the sources of radio bright patches (Section \ref{sec:pbp}).} 

\begin{figure}[h]
\centerline{ {\includegraphics[width=9cm]{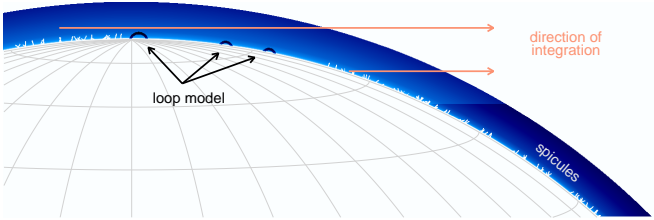}}}
\caption{The SSC atmospheric model with the presence of spicules and three small magnetic loops located in the spicules-less region. The arrows indicate the direction of the radiative transference integration at the polar region.}
\label{fig:model}
\end{figure}

Assuming that the NoRH maps have an spatial resolution of 10'', the SSC model without the inclusion of spicules  showed a limb brightening of 36\% above the quiet Sun values, which is compatible with the maximum values observed at the poles. The inclusion of spicules reduced the initial limb brightening to $\sim10\%$, that is close to the values observed at equatorial regions.

{  \cite{Selhorst2005b} explain the high polar brightening values by }the presence of holes in the spicule forest caused by intense magnetic features (i.e, faculae){  , hereafter referred to as bare regions}. For a large bare region located between $80.2-90.0^\circ$ heliographic angle,  the simulation results showed a sharp and intense brightening, up to 40\% above the quiet Sun. However, the intensity decreased to $21.4\%$ for a low latitude bare region located between $76.2-81.4^\circ$. Thus, the authors concluded that a simple hole in the spicule forest is only able to reproduce the brightness temperature increase caused by large bright patches very close to the limb ($\gtrsim 80^\circ$).    

 {However, since} the high intensity bright patches at 17~GHz were also observed at lower heliographic angles, another physical source is necessary to explain their brightness temperature values. As first suggested in \cite{Selhorst2010}, the simulations presented here include small magnetic loops in the regions without spicules, {  within} coronal holes.

\subsection{Coronal holes}\label{sec:ch}

\begin{figure}[h]
\centerline{ {\includegraphics[width=9cm]{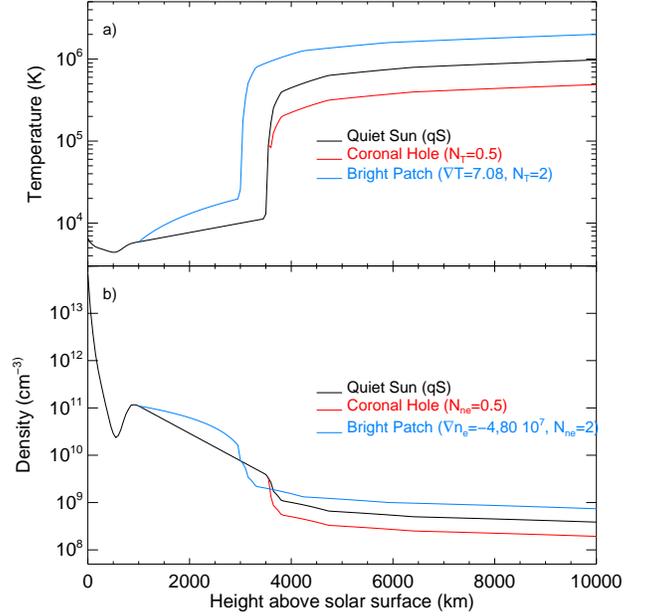}}}
\caption{(a) Temperature and (b) electron density  {profile} 
 for the quiet Sun (black curves), coronal holes (red curves) and bright patches (blue curves).}
\label{fig:SSC}
\end{figure}

{  Using the SSC models as a starting point,} the presence of coronal holes was simulated by reducing the coronal temperature and density (electron and ions) distributions, above 3650~km, in which the original values were multiplied by constant values, $N_T$ and $N_{ne}$. Figure~\ref{fig:SSC} shows (a) temperature and (b) electron density  {profiles} 
for the quiet Sun (black curves), coronal holes (red curves) and  bright patches (blue curves, see Section \ref{sec:pbp}). 

\begin{figure}[h]
\centerline{ {\includegraphics[width=9cm]{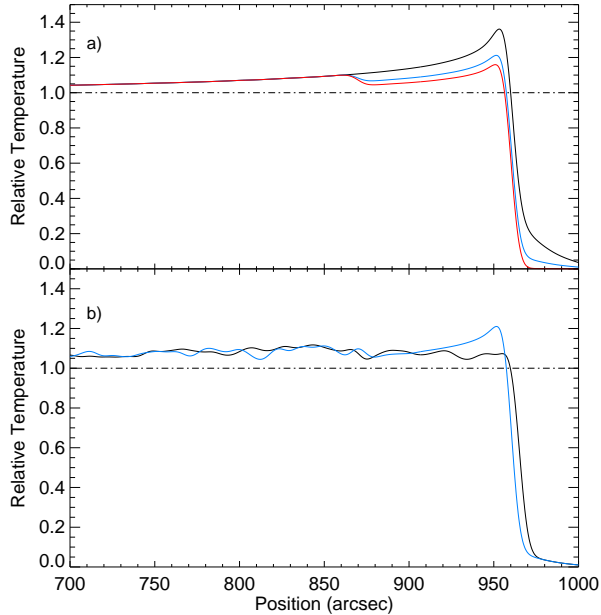}}}
\caption{a) Center-to-limb brightness temperature variation for the standard SSC model (black curve) and for a simulation with a coronal hole located at $65^\circ$ of latitude. {  The blue and red curves were obtained, respectively, for plumes and inter-plumes temperature and density distributions}. b) The influence of the inclusion spicules in the simulation with coronal hole { (plumes region)}. The black curve resulted from simulations with spicules distributed throughout the limb, while the blue one considered a spicule-less region above $70^\circ$ of latitude. 
}
\label{fig:CH}
\end{figure}

The temperature reduction in the coronal hole was set {  as $N_T=0.5$, whereas the densities were reduced by a factor $N_{ne}=0.5$. These settings resulted in temperatures and densities of $0.49\times10^6$~K and $1.93\times10^8\rm cm^{-3}$ at $10$~Mm above the surface, while, at $40$~Mm the values were $0.71\times10^6$~K and $0.79\times10^8\rm~cm^{-3}$. The upper atmosphere temperature and density are compatible with plumes regions reported by \cite{Wilhelm2006}}. The simulation of a coronal hole located at $65^\circ$ of latitude with the characteristics above, present a reduction in the 17~GHz limb brightening (21\%) when compared with the standard SSC simulation without spicules (36\%), this comparison is showed in Figure~\ref{fig:CH}a. {  If the inter-plume  temperature and density, observed by \cite{Wilhelm2006}, were adopted ($N_T=0.8$ and $N_{ne}=0.1$), the limb brightening was reduced to 16\%.} All simulations in this work were convolved with a $10''$ Gaussian beam, that represents the NoRH best spatial resolution. 

Similar to the procedure used in  previous works \citep{Selhorst2005a,Selhorst2005b} to estimate the contribution of 
spicules in coronal holes, they were randomly distributed in the temperature and densities matrices covering about  $10\%$ of the solar surface. Except for the width, that was fixed at 500~km, all spicules physical parameters were randomly chosen, with temperatures ranging from 7,000 to 13,000~K, densities in the interval of $2-6\times 10^{10}\rm$, heights from 5,000 to 7,000~km and  inclination angles from $30$ to $150^\circ$. {  These parameters are consistent with the values inferred from observations of optical lines, mainly $\rm H\alpha$ and $\rm Ca~II$ \citep{Sterling2000,Tsiropoula2012}}.

The brightness temperature was calculated every 100~km, instead of the 700~km used in previous works. To obtain a final mean profile,  $N$ simulations were performed until a convergence criterion was satisfied, that is, the rms of the 400 points of the mean profile closest to disk center should be less than 0.0003 in comparison with the rms calculated in the previous simulation. Usually, it sufficed to perform $20-40$ simulations.   

As can be seem in Figure~\ref{fig:CH}b (black curve), the presence of spicules prevent the {  reduction of the} brightness temperature caused by the coronal hole at low latitude angles ($\lesssim 930'' $ in Figure~~\ref{fig:CH}b). Moreover, the limb brightening is completely absorbed by the presence of spicules, in agreement with \cite{Selhorst2005a,Selhorst2005b}. As a result, although the inclusion of spicules in a plume region produces an emission at the limb 7\% more intense than the quiet Sun, this brightening cannot be {  identified from the emission originating in their surroundings.}

As has been proposed by \cite{Selhorst2005b}, the intense limb brightening can be caused by the magnetized regions that inhibit the presence of spicules causing a hole in the spicules forest. Nevertheless, a polar bare region within a coronal hole cannot reproduce the high 17~GHz bright patches temperatures. The blue curve in Figure~\ref{fig:CH}b differs from the black one by the inclusion of a polar bare region above $70^\circ$ of latitude.

{  The simulated spicules are optically thick at 17GHz, and their adopted density range lies around the lowest density values reported \citep{Tsiropoula2012}. This absorption is caused by the optical thickness of the spicules at 17~GHz, which, in the SSC model, is formed ($\tau\sim1$) around a region 2,900~km above the solar surface, where the local density and temperature are $9.3 \times 10^9\rm cm^{-3}$ and $10,390$~K, respectively. Since, this density is approximately half of the {minimum density value adopted for the} spicules, all the spicules reaching heights above 2,900~km are optically thick at 17~GHz.}

{  Moreover, observationally, spicules are not easily identified in the chromosphere, being predominantly seen when reaching above chromospheric heights \citep{Pereira2014}. Recent simulations suggest that spicules do not maintain their structure in the chromosphere, but only {\em may become} spicules once the chromospheric material flows upwards along the magnetic field strands \citep{Martinez2017}. For these reasons, we designed our spicules model to focus on the main aspects affecting the formation/propagation of the radio emission.} 

\subsection{Polar bright patches}\label{sec:pbp}

Since the spicules-less regions below $~80^\circ$ of latitude are not able to reproduce the intense bright patches observed in the NoRH maps \citep{Selhorst2005b} and the presence of coronal holes reduce the expected limb brightening at 17~GHz (Figure~\ref{fig:CH}),  other solar features should be acting inside the coronal holes to increase their brightness temperature at 17~GHz \citep{Gopal1999,Oliveira2016}. To simulate the observed 17~GHz bright patches, we introduced small magnetic loops inside coronal hole regions (Figure~\ref{fig:model}). {  In these simulations the coronal hole was set to have temperature and density distributions consistent with plume regions (red curves in Figure~\ref{fig:SSC}).} 
            
The simulated magnetic loops  were set as half circumference, perpendicular to the solar surface, with two possible external radius of 5.0~Mm (6.9'') and 7.5~Mm (10.3'') and fixed  {cross-section} width of 2.5~Mm. Inside the magnetic loop the temperatures and densities varied as an active region flux tube \citep{Selhorst2008}, i.e., hotter and denser than the atmosphere surrounding it. The blue curves in Figure~\ref{fig:SSC} show the example of the assumed atmospheric variation in the magnetic loops, in which the chromospheric gradients of temperature and density were $\nabla T=7.08~K~km^{-1}$ and $\nabla n_e=-4.08\cdot 10^7~cm^{-3}~km^{-1}  $, respectively, the transition region was considered to be at 3,000~km. Moreover, the coronal temperature and densities were considered as twice the quiet atmospheric values.

Table~\ref{table1} lists 30 bright patches simulations, with distinct plasma compositions and locations. The reference simulation numbers are placed in the first column, with different distribution of temperature ($\nabla T$ and $N_T$) and density ($\nabla n_e$ and $N_{n_e}$) within the flux tubes organised in the next four columns, followed by the magnetic loops size and position. The simulations outcomes are listed in the last four columns, where the first two are the maximum brightness temperature, $T_{B_{max}}$, and the width obtained at the point in which $T_B=0.5T_{B_{max}}$ obtained from the bright patch simulation without the convolution with the NoRH beam, whereas the last two represent the same values after the beam convolution. All small magnetic field loops were simulated inside the coronal hole limits. The widths were measured at half power of the maximum brightness temperature of the bright patches, after the subtracting the coronal hole brightness temperature profile (blue curve in Figure~\ref{fig:CH}a).                

\section{Results}

\begin{figure}[h]
\centerline{ {\includegraphics[width=9cm]{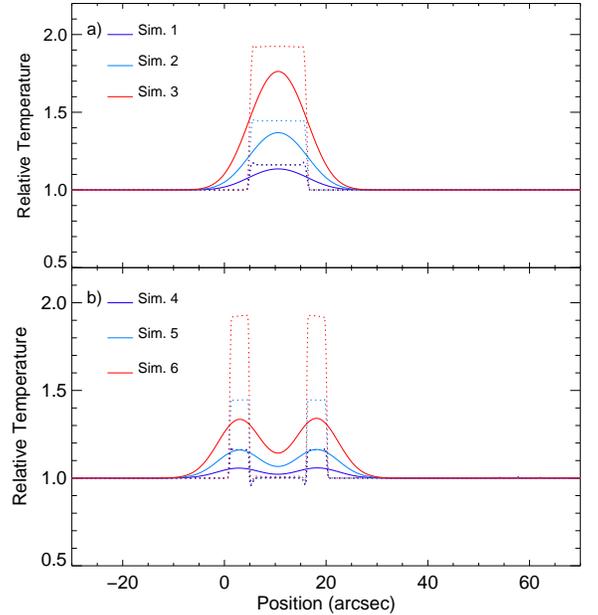}}}
\caption{Simulations of magnetic field loops located at the center of solar disc (1--6 on Table~\ref{table1}). Dotted lines are the unconvolved results, while continuous lines are the convolved ones.}
\label{fig:arc}
\end{figure}

Since most of the 17~GHz emission is generated in the chromosphere, the size of the magnetic loop determines the position where the emission is produced. While in the smaller loops (5.0~Mm) the emission is formed at the loop top, the emission in the larger loops (7.5~Mm) comes from their footpoints, whereas their tops are optically thin. In the first six simulations presented in Table~\ref{table1} the loops are placed at the center of the solar disk.  These results show the brightness temperature increase due to the larger gradients of the chromospheric temperatures and densities. In these simulations, when the plasma composition inside the loop is the same,  the obtained $T_{B_{max}}$ is independent of the magnetic loop size before the beam convolution, {  as expected}. Nevertheless, smaller loops {  produced a bright patch ($11''.3$) larger than the larger loops ($7''.9$)}. {  This is easily explained by the different sizes of their emitting areas: while for the smaller loops the loop top is bright, for the larger loops only the footpoints are brighter than the surroundings.} {  This can be visualized in} Figure~\ref{fig:arc}, which shows the results of simulations using (a) 5.0~Mm and (b) 7.5~Mm loops. The dotted lines are the {  unconvolved} results with 100~km spatial resolution and the continuous lines are the result of the convolution with the NoRH beam ($\sim 10''$).  Because the emitting area of the footpoint is smaller than the NoRH resolution, after the beam convolution, $T_{B_{max}}$ reduces more significantly in the brighter loop (Figure~\ref{fig:arc}b). Moreover, the convolved brightness temperature profile still presents a double peak with $\sim 21''$ width, i.e. more than $10''$ increase in width. On the other hand, the small loop (Figure~\ref{fig:arc}a) shows a smaller reduction in $T_{B_{max}}$, a single peaked profile and less than $2''$ increase in its width.           

\begin{figure}[h]
\centerline{ {\includegraphics[width=9cm]{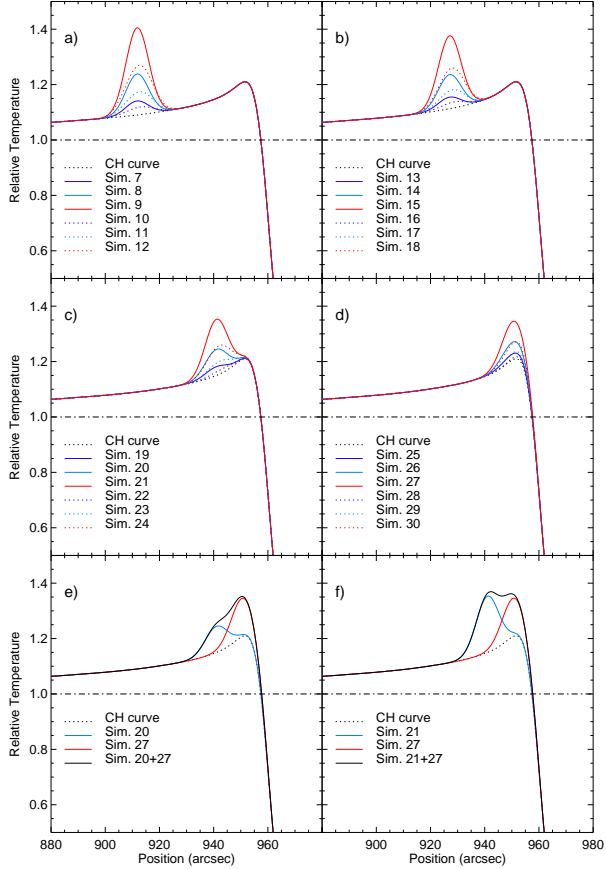}}}
\caption{Simulations of bright patches located at a) $71.1^\circ$, b) $74.1^\circ$, c) $77.4^\circ$ and d) $81.4^\circ$. Different colours were used to refer to the distinct simulations listed in Table 1. In panels e) and f) were simulated two bright patches located in different longitudes, respectively,  $77.4^\circ$ and $81.4^\circ$.}
\label{fig:BP}
\end{figure}

Due to {  projection effects and the curvature of the Sun}, the size of the emitting magnetic loops is {strongly} reduced when they are simulated at high latitudes. The unconvolved width of the hotter small loops was reduced from $\sim 11''$  at disk center to $\sim 1''$ when they are placed at $81.4^\circ$ (Sim. 3 and 25 in Table~\ref{table1}). After the convolution the resulting bright patches mimic the beam size ($\sim10''$). With respect to their brightness temperatures, $T_{B_{max}}$ is seen to increase with the angular position in the unconvolved values, however, due to the reduction in the emitting source size, the convolved values follows in the opposite way, reducing the $T_{B_{max}}$ values. 

The profile of simulations 7 to 30 from Table~\ref{table1} are plotted in Figure~\ref{fig:BP}a, b, c and d. The continuous lines refer to the small magnetic loops and the dotted ones represent the larger loops, for the same plasma configuration (temperature and electron density) the same color is used. {  The emission from loops located at $71.1^\circ$ and $74.1^\circ$ can be identified apart from} the limb brightening (Figure~\ref{fig:BP}a and \ref{fig:BP}b). However, those located at $77.4^\circ$ and $81.4^\circ$ cannot be distinguished from the {  usual} limb brightening (Figure~\ref{fig:BP}c and \ref{fig:BP}d). 

{  We also tested the effects of including two magnetic loops, using the physical conditions of simulations 20 and 27 (Figure~\ref{fig:BP}e) and simulations 21 and 27 (Figure~\ref{fig:BP}f.)} {  Note that the $T_B$ increase caused by each loop cannot be resolved after the convolution with the NoRH beam. Although not shown in the figure, for the unconvolved results with 100~km resolution, the $T_B$ increase caused by each loop can be easily resolved}. {  Here}, $T_{B_{max}}$ in both simulations is the same as that of simulation 27; {  however,} the width increased to $\sim19''.7$  and $21''.1$ in the profiles plotted in Figure~\ref{fig:BP}e and \ref{fig:BP}f, respectively.  

\section{Discussion and conclusions}          

The purpose of this work is to model the emission of the 17~GHz polar bright patches, which are frequently observed in the NoRH maps in association with coronal holes \citep{Gopal1999,Selhorst2003,Oliveira2016}. The simulations were based on the temperature and density distributions proposed in the SSC atmospheric model \citep{Selhorst2005a}, {  with modifications to include a coronal hole atmospheric model and magnetic loops as the sources of the radio bright patches.}

{  We have calculated the radio emission at 17~GHz from coronal holes, in comparison with typical quiet Sun regions. As expected, in a static atmosphere, the lower temperature and density (red profiles in Figure~\ref{fig:SSC}) inside a coronal hole resulted in lower brightness temperature values} (Figure~\ref{fig:CH}a).
{  Our results show, however, that the presence of (spatially unresolved) spicules can produce brighter regions than what would be expected from coronal holes (Figure~\ref{fig:CH}b).}

To simulate the bright patches, {  we have introduced small magnetic loops, with hotter and denser plasma than its surroundings. We find that the radio emission from smaller loops (5.0~Mm of radius) comes from the top of the loop, while the emission from larger loops (7.5~Mm of radius) originates from the footpoints.} As a consequence, the size of the {  simulated bright patches originating from} small loops was larger than {  from} larger loops, $\sim11''$ and $\sim8''$, respectively. However, after convolving the results with NoRH beam, {  large loop produced} broader and colder bright patches in comparison with the results obtained for the small loops.

{  The inclusion of magnetic loops in the model only affect the radio brightness at 17~GHz if their temperature and density properties are substantially different from the surrounding plasma at heights where $\tau \approx 1$, which happens near the upper chromosphere and lower corona. These results are in agreement with the findings of \cite{Brajsa2007}. Moreover,  loop models with increased density and temperature only at coronal heights do not contribute significantly to the radio emission at 17~GHz. Such loops could be brighter at EUV wavelengths, and thus} this could explain the absence of a one-to-one correlation between the 17~GHz bright patches and those observed in EUV \citep{Nindos1999,Riehokainen2001,Nitta2014}.


{  The maximum brightness temperature} $T_{B_{max}}$ in the simulations increased up to $~30\%$ {  by placing the loops at higher latitudes}, however, the size of these bright regions were reduced to $~1''$, {  much smaller than the NoRH beam}. As a consequence, after the beam convolution, the size of the bright patch corresponds to the beam size {  (as expected)}, in agreement with the minimum size of the bright patches observed in the NoRH maps \citep{Nindos1999}.    

On the other hand, a single small loop located in near the pole is not able to reproduce the  larger bright patches sizes observed ($50''-55''$, \cite{Nindos1999}), which could be caused by the presence of a great number of small magnetic loops unresolved by the NoRH beam. As showed in Figure~\ref{fig:BP}e and \ref{fig:BP}f, even loops separated  by an angular distance of $3^\circ$  ($\sim 37$~Mm) will not be resolved in the NoRH maps. Moreover, the presence of small magnetic loops close to the limb can result in a merged brightness limb profile, increasing the observed limb brightening temperature and width.

To improve our knowledge about these small bright structures inside the coronal holes, high resolution observations at different wavelengths are necessary. Today, only solar observations with ALMA can achieve these spatial resolution \citep{Wedemeyer2016}.

\startlongtable 
\begin{table*}[ht]
\begin{center}
\begin{footnotesize}
\renewcommand{\arraystretch}{0.5}
\caption{{The 17~GHz bright patches simulations and their free parameters.}\label{table1}}
\begin{tabular}{c c c c c c c | c c c c}
\hline \hline
&{$\nabla T$}&{$N_T$}&{$\nabla n_e$}&{$N_{n_e}$}& Loop Radius &Loop Position & \multicolumn{4}{c}{Bright Patches Results}\\
&{ (K~km$^{-1}$)}& & { (cm$^{-3}$~km$^{-1}$)}& &{ (Mm)} & {$ (^\circ)$ }&  \multicolumn{2}{c}{unconvolved}&  \multicolumn{2}{c}{Convolved }\\
& & & & & & &$T_{Bmax}~(\times 10^3)$~K & Width ('') &$T_{Bmax}~(\times 10^3)$~K & Width ('')   \\
\hline
1&3.04&1.2&$-5.14\cdot 10^7$&1.2&5.0& 0 & 12.2 (T) & 11.3 & 11.8 & 13.1\\
2&4.56&1.5&$-5.01\cdot 10^7$&1.5&5.0& 0 & 15.1 (T)& 11.3 & 14.2 & 13.0\\
3&7.08&2.0&$-4.80\cdot 10^7$&2.0&5.0& 0 & 20.0 (T)& 11.3  & 18.3 & 13.0\\
4&3.04&1.2&$-5.14\cdot 10^7$&1.2&7.5& 0 & 12.2 (F)& 7.9 & 11.0 & 20.7\\
5&4.56&1.5&$-5.01\cdot 10^7$&1.5&7.5& 0 & 15.1 (F)& 7.9  & 12.1 & 20.7\\
6&7.08&2.0&$-4.80\cdot 10^7$&2.0&7.5& 0 & 20.0 (F)& 7.7 & 13.9  & 21.2\\
7&3.04&1.2&$-5.14\cdot 10^7$&1.2&5.0& 71.1 & 13.3 (T) & 3.2 & 11.9 & 9.9\\
8&4.56&1.5&$-5.01\cdot 10^7$&1.5&5.0& 71.1 &  16.6 (T) & 3.2  & 12.9 & 9.9 \\
9&7.08&2.0&$-4.80\cdot 10^7$&2.0&5.0& 71.1 & 22.5 (T) & 3.2 & 14.6 & 9.9\\
10&3.04&1.2&$-5.14\cdot 10^7$&1.2&7.5& 71.1 & 13.3 (F) & 1.9 & 11.6 & 11.7\\
11&4.56&1.5&$-5.01\cdot 10^7$&1.5&7.5 & 71.1 &16.6 (F) & 1.8 & 12.2 & 11.7\\ 
12&7.08&2.0&$-4.80\cdot 10^7$&2.0&7.5 & 71.1 &  22.5 (F) & 1.9 & 13.2 & 11.6\\
13&3.04&1.2&$-5.14\cdot 10^7$&1.2&5.0&74.1 & 13.8 (T) & 2.5 & 12.6 & 9.8\\
14&4.56&1.5&$-5.01\cdot 10^7$&1.5&5.0&74.1 & 16.9 (T) &2.6& 12.8 &9.8 \\
15&7.08&2.0&$-4.80\cdot 10^7$&2.0&5.0&74.1& 22.8 (T) & 2.6 & 14.3 & 9.8 \\
16&3.04&1.2&$-5.14\cdot 10^7$&1.2&7.5&74.1 & 13.6 (F) & 1.4 & 12.6 & 11.0\\
17&4.56&1.5&$-5.01\cdot 10^7$&1.5&7.5&74.1 & 16.9 (F)&1.4& 12.6 &11.0 \\
18&7.08&2.0&$-4.80\cdot 10^7$&2.0&7.5&74.1& 22.9 (F) & 1.4 & 13.1 & 11.0 \\
19&3.04&1.2&$-5.14\cdot 10^7$&1.2&5.0&77.4 & 14.1 (T) & 1.8 & 12.6 & 9.8\\
20&4.56&1.5&$-5.01\cdot 10^7$&1.5&5.0&77.4 & 17.7 (T) &1.8 & 12.9 &9.7 \\
21&7.08&2.0&$-4.80\cdot 10^7$&2.0&5.0&77.4& 23.8 (T) & 1.8 & 14.1 & 9.8 \\
22&3.04&1.2&$-5.14\cdot 10^7$&1.2&7.5&77.4 & 14.2 (F) & 0.8 & 12.6 & 10.5\\
23&4.56&1.5&$-5.01\cdot 10^7$&1.5&7.5&77.4 & 17.8 (F)&0.7& 12.6 &10.5 \\
24&7.08&2.0&$-4.80\cdot 10^7$&2.0&7.5&77.4 & 23.9 (F) & 0.7 & 13.1 & 10.6 \\
25&3.04&1.2&$-5.14\cdot 10^7$&1.2&5.0&81.4 & 21.5 (T) & 1.0 & 12.9 & 9.8\\
26&4.56&1.5&$-5.01\cdot 10^7$&1.5&5.0&81.4 & 21.5 (T) &1.1 & 13.2 &9.8 \\
27&7.08&2.0&$-4.80\cdot 10^7$&2.0&5.0&81.4& 25.3 (T) & 1.1 & 14.0 & 9.7 \\
28&3.04&1.2&$-5.14\cdot 10^7$&1.2&7.5&81.4 & 21.5 (F) & 0.4 & 12.6 & 9.9\\
29&4.56&1.5&$-5.01\cdot 10^7$&1.5&7.5&81.4 & 21.5 (F) &0.3 & 12.8&10.5 \\
30&7.08&2.0&$-4.80\cdot 10^7$&2.0&7.5&81.4& 23.4 (F) & 1.3 & 13.2 & 10.3 \\
\hline
\end{tabular}
\end{footnotesize}
\end{center}
\end{table*}

\acknowledgments
We would like to thank the Nobeyama Radioheliograph, which is operated by the NAOJ/Nobeyama Solar Radio Observatory. A.J.O.S. acknowledge the scholarship form CAPES. C.L.S. acknowledge financial support from the S\~ao Paulo Research Foundation (FAPESP), grant number 2014/10489-0. P.J.A.S. acknowledges support from grant ST/L000741/1 made by the UK's Science and Technology Facilities Council and from the University of Glasgow's Lord Kelvin Adam Smith Leadership Fellowship.



\end{document}